\documentclass[11pt,twoside]{article}
\usepackage{macro-fsut-eng}
\usepackage{graphicx}
\usepackage[T1]{fontenc}
\usepackage{latexsym}
\usepackage{verbatim}

\begin{document}

\vskip 1.0cm
\markboth{R.E.G.~Machado \& G.B.~Lima~Neto}{Simulations of galaxy cluster mergers}
\pagestyle{myheadings}

\vspace*{0.5cm}
\title{Simulations of galaxy cluster mergers: the dynamics of Abell 3376}

\author{Rubens E. G. Machado \& Gast\~ao B. Lima Neto}
\affil{Instituto de Astronomia, Geof\'isica e Ci\^encias Atmosf\'ericas, Universidade de S\~ao Paulo, R. do Mat\~ao 1226, 05508-090 \\S\~ao Paulo, Brazil}

\begin{abstract}
In large scale structure formation, massive systems assemble through the hierarchical merging of less massive ones. Galaxy clusters, being the most massive and thus the most recent collapsed structures, still grow by accreting smaller clusters and groups. In order to investigate the dynamical evolution of the intracluster medium, we perform a set of adiabatic hydrodynamical simulations of binary cluster mergers. 
\end{abstract}

~ \\

\noindent Abell 3376 is a nearby (z=0.046) rich galaxy cluster whose bullet-shaped X-ray emission suggests that it is undergoing a major collision, approximately on the plane of the sky. It has an estimated virial mass of $5.2 \times 10^{14} M_{\odot}$. This is the closest galaxy cluster with such morphology, to our knowledge. The brightest cluster galaxy lies far from the region of largest X-ray emission, a feature which must be accounted for by the reconstruction of its dynamycal history. Additionally, this cluster presents diffuse radio emission (due to the acceleration of relativistic electrons), whose morphology and intensity are compatible with a recent merger scenario (Bagchi et al. 2006).

With our simulations, using the parallel SPH code Gadget-2 (Springel 2005), we are able to obtain gas emissivity maps, comparable to those derived from X-ray observations. In particular, we attempt to model the intracluster gas morphology. Exploring the parameter space of initial conditions (specially mass ratios, impact parameters and relative velocity) allows us to set constraints on the original masses (baryonic and non-baryonic) of the colliding clusters and their dynamical history.

Our set of preliminary simulations suggests an approximately head-on collision with mass ratio of about 3:1. Simulated X-ray emission presents a morphology roughly comparable to that of Abell 3376 at approximately 0.2 Gyr after the instant of central passage. Projections angles larger than $10^{\circ}$ appear to give poorer results, indicating that the collision axis must lie close to the plane of the sky.

\acknowledgments This work was partially supported by FAPESP (2010/ 12277-9) and by the CAPES/COFECUB cooperation.

\end{document}